# General Graph Identification By Hashing

Tom Portegys
*Email: portegys@gmail.com*

## Abstract

A method for identifying graphs using MD5 hashing is presented. This allows fast graph equality comparisons and can also be used to facilitate graph isomorphism testing. The graphs can be labeled or unlabeled. The method identifies vertices by hashing the graph configuration in their neighborhoods. With each vertex hashed, the entire graph can be identified by hashing the vertex hashes.

## Introduction

Numerous uses could be found for unique and concise graph identifiers: graphs could then be counted, sorted, compared and verified more easily. For example, chemical compounds could be specified by identifying their constituent molecules represented by graphs of spacial and bonding relationships between atoms.

The problem with developing a method for identifying graphs is that graphs are very general objects. Uniquely identifying vertices and edges solves the problem but begs the question, since the problem then becomes how to arrive at these identifiers in a uniform fashion (Sayers and Karp, 2004).

The method proposed here identifies vertices by computing an MD5 hash (Rivest, 1992) for the graph configuration in their neighborhoods in an iteratively expanding manner. Once each vertex has a unique hash or is fully expanded, the vertex hashes are sorted and hashed to yield a hash for the graph, a technique similar to that used by Melnik and Dunham (2001).

## Description

### Graph format
Using a pseudo-C++ notation, the following define a graph vertex and edge:

```
Vertex
{
    int label;
    Edge edges[];
};

Edge
{
    int label;
    Vertex source;
    Vertex target;
    bool directed;
```

};

This general scheme allows for a number of graph variations: labeled/unlabeled (using null labels), directed/undirected (for undirected, source and target are synonymous), and multigraphs. An edge is stored with both vertices; but only once when an edge connects a vertex to itself. A set of vertices and edges comprises a graph, which may be connected or not.

### *Algorithm*

The following object is used to construct MD5 hash codes based on vertex graph neighborhoods:

```
VertexCoder
{
    Vertex vertex;
    Edge parentEdge;
    unsigned char code[MD5_SIZE];
    VertexCoder children[];
    VertexCoder creator;
    bool expanded;
    int generation;
    hashmap vertexMap[(Vertex, Edge) => VertexCoder];
    void generateCode(bool);
    void expand();
};
```

The algorithm iteratively expands each vertex in the graph into an acyclic directed subgraph of coder objects representing the vertices and edges in its neighborhood until each vertex either has a distinct code or is fully expanded, implying that there may be vertices with identical codes. The graph hash code is then constructed by sorting and hashing the vertex codes.

```
// Generate code.
// The boolean argument allows labels to be included in the
// hash calculation.
void generateCode(bool hashLabels)
{
    // The graph (root) coder has a null vertex and a child
    // for each vertex.
    if (vertex == null)
    {
        // Iteratively expand children and generate
        // codes without labels to encode structure.
        for (i = 0; i < children.length; i++)
        {
            children[i].expand();
            children[i].generateCode(false);
```

```
            }

            // Continue until fully expanded or all hashes
            // are distinct.
            while (true)
            {
                // All vertices are fully expanded?
                for (i = 0; i < children.length; i++)
                {
                    if (!children[i].expanded) break;
                }
                if (i == children.length) break;

                // All vertices are distinct?
                sort(children);
                for (i = 0; i < children.length - 1; i++)
                {
                    if ((!children[i].expanded ||
                         !children[i + 1].expanded) &&
                        children[i] == children[i + 1]) break;
                }
                if (i == children.length - 1) break;

                // Expand and generate codes for range of
                // equivalent children.
                for (j = i + 1; j < children.length &&
                     children[i] == children[j]; j++) {}
                for (; i < j; i++)
                {
                    if (!children[i].expanded)
                    {
                        children[i].expand();
                        children[i].generateCode(false);
                    }
                }
            }
        }

        // Recursively generate code.
        oldCode = code;
        for (i = 0; i < children.length; i++)
        {
            children[i].generateCode(hashLabels);
        }
        sort(children);
```

```
        // Construct code from vertex, parent edge, and
        // childrens' codes.
        if (vertex != null)
        {
            if (hashLabels) content += vertex.label;
            if (parentEdge != null)
            {
                if (hashLabels) content += parentEdge.label;

                // Codify the parent edge type.
                if (parentEdge.directed)
                {
                    if (parentEdge.source == vertex)
                    {
                        content += 0;
                    }
                    else
                    {
                        content += 1;
                    }
                }
                else   // Undirected edge.
                {
                    content += 2;
                }
            }
        }
        for (i = 0; i < children.length; i++)
        {
            content += children[i].code;
        }

        // Create code by hashing the content.
        code = md5(content);

        // Mark vertex as expanded if code not changed.
        if (oldCode == code) expanded = true;
    }

    // Expand coder one level deeper.
    void expand()
    {
        if (expanded) return;
        if (children.length == 0)
        {
            // Expand this vertex.
            for (i = 0; i < vertex.edges.length; i++)
```

```
            {
                if (vertex == vertex.edges[i].source)
                {
                    child =
                        vertexMap.find(vertex.edges[i].target,
                        vertex.edges[i]);
                    if (child == null)
                    {
                        // Create a new child coder:
                        // vertex = edge target, parent edge =
                        // edge, creator = this coder,
                        // generation + 1.
                        child =
                            new VertexCoder(
                                vertex.edges[i].target,
                                vertex.edges[i], this,
                                generation + 1);
                        children.append(child);
                        vertexMap[vertex.edges[i].target,
                            vertex.edges[i]] = child;
                    }
                    else
                    {
                        // Share next generation coder.
                        if (child.generation == generation + 1)
                        {
                            children.append(child);
                        }
                    }
                }
                else   // Vertex is target.
                {
                    child =
                        vertexMap.find(vertex.edges[i].source,
                        vertex.edges[i]);
                    if (child == null)
                    {
                        // Create a new child coder:
                        // vertex = edge source, parent edge =
                        // edge, creator = this coder,
                        // generation + 1.
                        child =
                            new VertexCoder(
                                vertex.edges[i].source,
                                vertex.edges[i], this,
                                generation + 1);
                        children.append(child);
```

```
                    vertexMap[vertex.edges[i].source,
                        vertex.edges[i]] = child;
                }
                else
                {
                    // Share next generation coder.
                    if (child.generation == generation + 1)
                    {
                        children.append(child);
                    }
                }
            }
        }
    }
    else
    {
        // Expand deeper.
        for (i = 0; i < children.length; i++)
        {
            // Only coder creator can expand to prevent
            // duplicate expansion.
            if (children[i].creator == this)
                children[i].expand();
        }
    }
}
```

### *Graph hash example*

The method is illustrated through an example. Consider the simple directed graph shown in Figure 1. The vertices and edges are labeled for illustrative purposes, but the algorithm works for unlabeled vertices and edges as well as undirected edges.

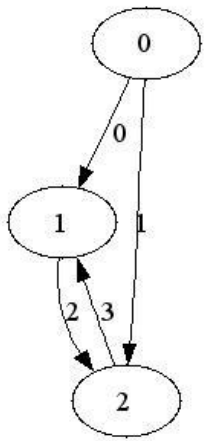

Figure 1 – A simple directed graph.

The hash coding algorithm will create the vertex coder graph shown in Figure 2 for the graph in Figure 1. The vertex coder graph is an acyclic directed graph (although the source graph need not be). The root of the vertex coder graph represents the entire source graph, and will contain the hash code for the source graph after processing is complete. The (f) and (b) notation on the edges represent a directed edge in the source graph in the forward and backward direction respectively.

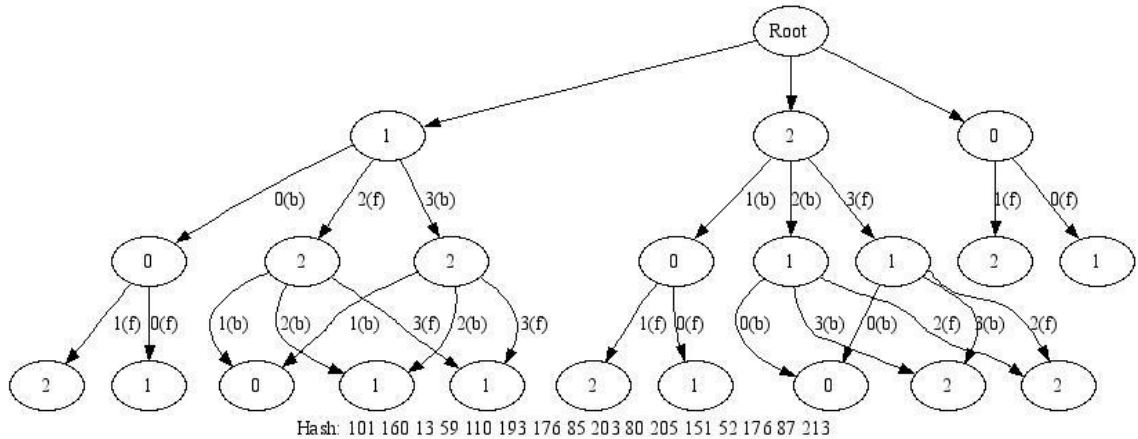

Figure 2 – Graph vertex coder

Before the first call to generateCode(), the coder is configured as in Figure 3. This configuration reveals nothing about the edges in the graph, and thus must always be expanded. Moreover, it should be noted that a graph is initially encoded without regard to vertex and edge labels. If the graph is labeled the labels are incorporated into the hash code in a final pass. The reason for this is that the algorithm terminates when the vertex codes are all unique; using labels could prematurely result in graphs that have differing edge configurations computing the same hash code. Each child of root is also initialized with its own map (*vertexMap*) used in subgraph expansion to detect and prevent cycles.

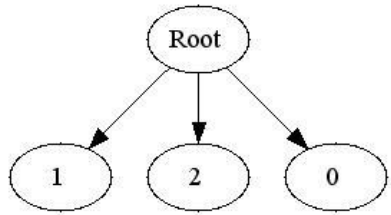

Figure 3 – Initial coder configuration.

After the first expansion, the coder appears as in Figure 4. Note that for vertex 0, there are 2 forward edges to vertices 1 and 2. For vertex 1, there is a forward and backward edge to vertex 2, and a backward edge to vertex 0. Vertex 2 has a forward and backward edge to vertex 1, and a backward edge to vertex 0. Considering that the algorithm is actually disregarding labels at this point, it can be seen that vertex 0's subgraph structure is distinct from that of vertex 1 and 2, both of which have identical structures. This fact is discovered after sorting and scanning the vertex hash codes, and for this reason vertex 1 and 2 must be expanded.

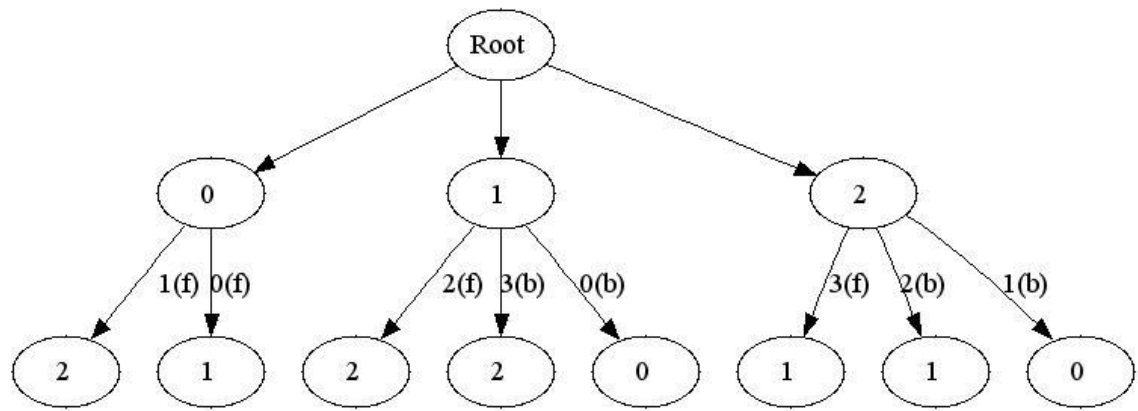

Figure 4 – First expansion.

Refer back to Figure 2 for the results of the next expansion of vertices 1 and 2. Several new coders become "shared" children, representing common states of expansion. What occurs during processing is that one parent initially creates the child, which is then detected by the *vertexMap* data structure during the expansion of the other parents. The *generation* member of the child distinguishes this as a valid child instead of a loop to a previous coder. In order to avoid duplicate expansions of a shared child, the only parent allowed to expand the child is the one indicated by the *creator* member of the child.

It can be seen that the subgraphs for these vertices are still structurally identical, which, looking at the source graph, is exactly the case. A subsequent call to expand() will result in no further subgraph expansion, since cycles will be detected by the *vertexMap* data structure. Because of this, the hash codes remain unchanged, marking the vertices as expanded and unavailable for further expansion.

### *Graph isomorphism*

The vertex coder graph can be used in determining isomorphism between labeled graphs, a task that is possibly NP-complete. Graphs are isomorphic if there is a consistent mapping between their labeled vertices. If the two candidate graphs are first hashed without regard to labels, the hash codes of the vertex coders can be used to constrain the vertex correspondences between the two graphs, since only those that have the same hash can possibly correspond.

## Performance

The performance of the algorithm was measured with a graph comparison task against the Ullman (1976) and Schmidt-Druffel (1976) algorithms. The Ullmann algorithm is one of the most commonly used for graph isomorphism because of its generality and effectiveness (Cordella, et. al., 2001). The results are shown in Table 1. For each setting of vertices/edges from 10/100 to 100/10000, 1000 directed and unlabeled randomly connected graphs were generated. To compare, each graph was cloned and the ordering of the vertices and edges randomly scrambled in the clone to prevent artifactual timings;

so each pair was isomorphic but differently organized internally. The results are given as the average of the measured times in milliseconds to do the comparisons. For the hash algorithm, the time to hash both graphs and compare the hashes for equality was measured. The tests were conducted on a SUN UltraSPARC III workstation.

Table 1 – Graph comparison times (ms).

| Vertices/Edges | Ullmann | Schmidt-Druffel | Hash |
|---|---|---|---|
| 10/100 | 0.0 | 0.01 | 19.06 |
| 20/400 | 1.09 | 2.00 | 64.92 |
| 30/900 | 1.96 | 6.00 | 138.43 |
| 40/1600 | 3.84 | 12.13 | 238.93 |
| 50/2500 | 6.25 | 21.82 | 389.99 |
| 60/3600 | 9.53 | 34.40 | 576.34 |
| 70/4900 | 13.69 | 51.13 | 775.20 |
| 80/6400 | 18.99 | 72.23 | 1076.49 |
| 90/8100 | 25.67 | 98.26 | 1386.25 |
| 100/10000 | 33.62 | 129.45 | 1840.11 |

These timings can be used to estimate when the use of hashing is advantageous. For example, for categorizing graphs, a set of graphs is compared against a set of exemplars that grows as unmatched graphs are encountered. With hashing, the comparison can be done with a binary search or use of a hash space (hashing the graph hashes) to increase speed. For $N$ exemplars, and using a binary search, the tipping point can be expressed as follows:

$$\frac{Ullmann \times N}{2} = (\log_2(N) \times CMP) + Hash$$

Where:
*Ullmann* is Ullmann algorithm comparison time.
*Hash* is the time to hash a graph.
*CMP* is the time to compare MD5 hashes.

Using the data in Table 1, the hashing method is preferable when $N > 150$.

Additional measurements were taken to confirm an expected increase in hashing time for "symmetrical" graphs, i.e. graphs that have numerous structurally identical vertex coder subgraphs. As expected, this significantly increased the processing time as well as the memory needed to hold the larger subgraphs. For example, the time to hash a 100-vertex fully interconnected graph was 48.867 seconds.

## Conclusion

A method for identifying graphs using MD5 hashing has been presented. The algorithm grew out of a need for identifying molecules in a chemistry simulation constructed by the author. A simpler initial method was later elaborated into the one given here.

Measurements indicate that the method shows promise as an effective means of comparing graphs.

## References

L. P. Cordella, P. Foggia, C. Sansone, and M. Vento (2001), An Improved Algorithm for Matching Large Graphs, Proceedings of International Workshop on Graph-based Representation in Pattern Recognition, Ischia, Italy, pp. 149 - 159.

S. Melnik (2001), RDF API draft: Cryptographic digests of RDF models and statements, http://www-db.stanford.edu/~melnik/rdf/api.html#digest.

R. Rivest (1992), RFC 1321 The MD5 Message-Digest Algorithm, http://tools.ietf.org/html/rfc1321.

C. Sayers and A. H. Karp (2004), RDF Graph Digest Techniques and Potential Applications, Mobile and Media Systems Laboratory, HP Laboratories Palo Alto, HPL-2004-95.

D. C. Schmidt and L. E. Druffel (1976), A Fast Backtracking Algorithm to Test Directed Graphs for Isomorphism Using Distance Matrices, Journal of the Association for Computing Machinery, 23, pp. 433-445.

J. R. Ullmann (1976), An Algorithm for Subgraph Isomorphism, Journal of the Association for Computing Machinery, vol. 23, pp. 31-42.